\documentclass[11pt,a4paper]{article}
\begin{document}
\sloppy
\title{The standard model \\and the constituents of leptons and quarks}
\author{Walter Schmidt-Parzefall \\
Universit\"at Hamburg} 
\maketitle 
\begin{abstract}
A complete set of postulates of the standard model of the electroweak interaction and mass generation is formulated and confirmed deriving the Lagrangian for the standard model. A massive fermion is formed by a right-handed and a left-handed elementary massless fermion, exchanging a scalar doublet. The elementary massless fermions forming leptons belong to an approximate $SU(3)$ octet. The charges are quantised due to this symmetry. \\

Pacs 12.10 Dm 

Standard model; Higgs mechanism; electroweak interaction
\end{abstract} 
\vspace*{60pt}

The standard model of the electroweak interaction and mass generation, the model of Glashow, Weinberg and Salam [1], very successfully describes all experimental observations in particle physics made so far. The recent discovery of the Higgs particle at CERN by ATLAS and CMS is the final proof of the validity of the standard model. After the discoveries of neutrino oscillations and of the Higgs particle, the standard model deserves a detailed and consistent formulation. \\

Describing the minimal standard model in an unconventional language, in this note it is assumed that leptons and quarks are formed by particles, which are more elementary. From their properties a set of postulates of the minimal standard model is worked out. As a proof of the postulates, the well established Lagrangian of the standard model is derived. \\

For a definition of the notation, some essentials of Yang-Mills theory [2], on which the electroweak interaction is based, are provided in the Appendix. 

\pagebreak
In order to derive the Lagrangian for the electroweak interaction [3], it is assumed that a massive fermion doublet state $\psi_i$, like a neutrino for $i=1$ or an electron for $i=2$, is composed of constituents: A massless left-handed fermion doublet state $\chi_{iL}$ and a massless right-handed fermion singlet state $\psi_{iR}$ exchanging a scalar doublet $\phi$. These three primary elementary fields have to be chosen with the following properties: \\

1. The left-handed massless fermion doublet $\chi_L$ has $SU(2)\times U(1)$ symmetry and is represented by 
\begin{equation} \chi_L = {a \choose b} \frac{1 - \gamma^5}{2} \, \psi_i \, , \end{equation}
where $\psi_i = \psi_i \, (x^\mu)$ is a 4-spinor and the parameters $a$ and $b$ are complex numbers with
\begin{equation}  a^*a +b^*b = 1 \, . \end{equation}
According to (70), the $SU(2)\times U(1)$ interaction transformation $U_\chi$ for the interaction of $\chi_L$ with massless gauge potentials is given by
\begin{equation} U_\chi = \exp \Big(- \mbox{i} \int \left( \frac{g}{\sqrt{2}} (T_+ W_\mu^+ + T_- W_\mu^- \Big) + g \, T_3 W_\mu^3 + g^\prime Y B_\mu \right) \mbox{d} x^\mu \Big) . \end{equation} 
A constituent $\chi_{iL}$ of a massive fermion is an eigenstate of $T_3$, represented by 
\begin{equation} \chi_{iL} = {a \choose b}_{\! i} \frac{1 - \gamma^5}{2} \, \psi_i \qquad \mbox{with} \qquad {a \choose b}_{\! 1} = {1 \choose 0} \qquad \mbox{or} \qquad {a \choose b}_{\! 2} = {0 \choose 1} \, . 
\end{equation}
The interaction transformation for a state $\chi_{iL}$ is
\begin{equation} U_\chi = \exp \Big( - \mbox{i} \int \left( \frac{g}{\sqrt{2}} (T_+ W_\mu^+ + T_- W_\mu^- ) + g \, t_3 W_\mu^3 + g^\prime y B_\mu \right) \mbox{d} x^\mu \Big) , \end{equation}
where $t_3$ is the third component of the weak isospin and $y$ is the weak hypercharge of $\chi_{iL}$. The two states  $\chi_{iL}$ carry the same weak hypercharge $y$ but different weak isospin $t_3 = \pm \frac{1}{2}$. \\

2. The right-handed massless fermion singlet $\psi_{iR}$ has $U(1)$ symmetry and is represented by 
\begin{equation} \psi_{iR} = \frac{1 + \gamma^5}{2} \, \psi_i \, . \end{equation}
A state $\psi_{iR}$ carries zero weak isospin and the $U(1)$ charge $g^\prime y_i$. The $U(1)$ interaction transformation for a state $\psi_{iR}$ is according to (56)
\begin{equation} U_\psi = \exp \Big( - \mbox{i} \int g^\prime y_i \, B_\mu \, \mbox{d} x^\mu \Big) . \end{equation}
The two states  $\psi_{iR}$ carry the same weak isospin $t_3 = 0$ but different weak hypercharges $y_i$. \\

The states $\chi_{iL}$ and $\psi_{iR}$ with equal index $i$ together form a massive fermion $\psi_i$. \\

\pagebreak
The massless fermions $\chi_{iL}$ and $\psi_{iR}$ carry electric charge. The electric charge is a linear combination of the weak charge $g \, t_3$ and the $U(1)$ charge, defined as the linear combination for which both states $\chi_{iL}$ and $\psi_{iR}$, together forming a massive fermion $\psi_i$, are carrying the same electric charge $q_i$. \\

Expressing the interaction transformations $U_\chi$ and $U_\psi$ in terms of the electromagnetic potential $A_\mu$, belonging to the electric charge $q_i$, and the orthogonal potential $Z_\mu$, belonging to the generator $f$, results in
\begin{equation} g \, t_3 W_\mu^3 + g^\prime y \, B_\mu = f_\chi \, Z_\mu + q_i \, A_\mu \, , \qquad g^\prime y_i \, B_\mu = f_\psi \, Z_\mu + q_i \, A_\mu \, . \end{equation}
These equations are valid if the potentials and the generators are rotated by the same angle $\theta_W$ 
\begin{eqnarray} 
Z_\mu &=& \cos \theta_W W_\mu^3 - \sin \theta_W B_\mu \nonumber\\
A_\mu &=& \sin \theta_W W_\mu^3 + \cos \theta_W B_\mu \\[15pt]
f_\chi \! &=& \cos \theta_W g \, t_3 - \sin \theta_W g^\prime y \, , \qquad \; f_\psi \; = - \sin \theta_W g^\prime y_i \nonumber\\
q_i &=& \sin \theta_W g \, t_3 + \cos \theta_W g^\prime y \, \; \;  = \, \; \; \cos \theta_W g^\prime y_i \, . 
\end{eqnarray} 
The electric charge is thus defined. The electroweak angle $\theta_W$, the Weinberg angle, is given by
\begin{equation} \tan \theta_W = \frac{g^\prime y_i - g^\prime y}{g \, t_3} \, . \end{equation}

Expressing the generators $f$ in terms of $q_i$ results in
\begin{equation} f_\chi = \frac{g \, t_3}{\cos \theta_W} - \tan \theta_W \, q_i \, , \qquad \qquad f_\psi = - \tan \theta_W \, q_i  \, . \end{equation}
After replacing the generators $f$ in (8) by this result, the interaction transformations $U_\chi$ and $U_\psi$ are representing a unified description of the weak and electromagnetic interactions
\begin{eqnarray} U_\chi \!\!\!\! &=& \!\!\! \exp \Big( \! - \mbox{i} \! \int \! \left( \frac{g}{\sqrt{2}} (T_+ W_\mu^+ + T_- W_\mu^- ) + \frac{g \, t_3}{\cos \theta_W} \, Z_\mu + q_i \left( A_\mu - \tan \theta_W Z_\mu \right) \right) \mbox{d} x^\mu \Big) \nonumber\\[8pt]  
U_\psi \!\!\!\! &=& \!\!\! \exp \Big( - \mbox{i} \int q_i \left( A_\mu - \tan \theta_W Z_\mu \right) \mbox{d} x^\mu \Big) . \end{eqnarray} 
The charged gauge potentials $W_\mu^\pm$ carry the elementary electric charge $\pm e \, $. Due to conservation of electric charge the elementary electric charge $e$ is therefore
\begin{equation} e = q_1 - q_2 = g \, \sin \theta_W = g^\prime \cos \theta_W \end{equation}
where in addition the $U(1)$ unit charge $g^\prime$ has been defined for the convenience to write the electric charge $q_i$ given by (10) as a simple relation 
\begin{equation} q_i = e \left( t_3 + y \right) = e \, y_i \, . \end{equation} 

From the observed electric charge $q_i$ the weak hypercharge of the constituents is
\begin{equation} y = - \frac{1}{2}, \; y_1=0, \; y_2=-1 \; \mbox{for leptons,} \quad y = \frac{1}{6}, \; y_1=\frac{2}{3}, \; y_2=-\frac{1}{3} \; \mbox{for quarks} \, . \end{equation}
The experimental value of the electroweak angle is [3] $\theta_W \approx 28.6^\circ$. In order to see if there is a symmetry responsible for this value of $\theta_W$, it is instructive to plot the charges of the constituents of leptons, including the corresponding anti-particle states, and to add the electric charge as an index. For example the first generation of massive leptons is composed of the constituents
\begin{eqnarray} 
\nu_e \, : \quad \chi_{1L}^0 , \psi_{1R}^0 &\qquad& e^- \, : \quad \chi_{2L}^- , \psi_{2R}^- \nonumber\\[4pt]
\overline{\nu_e} \, : \quad \chi_{2R}^0 , \psi_{2L}^0 &\qquad& e^+ \, : \quad \chi_{1R}^+ , \psi_{1L}^+ \, .
\end{eqnarray}
 
\setlength{\unitlength}{1.1cm}
\begin{picture}(12,6.3)
\put(6,0.6){\vector(0,1){5}} \put(5.8,5.7){$g^\prime y$}
\put(3.5,3){\vector(1,0){5.1}} \put(8.7,2.93){$g\,t_3$}
\qbezier(4.73,0.8)(6,3)(7.27,5.2)
\put(7.1,4.93){\vector(2,3){0.2}} \put(7.36,5.3){$q$}
\qbezier(6.87,4.5)(6.92,4.47)(6.97,4.44) \put(6.64,4.45){$\scriptstyle e$}
\qbezier(5.14,1.5)(5.19,1.47)(5.24,1.44) \put(4.78,1.45){-$\scriptstyle e$}
\put(4.27,3){\line(0,1){0.12}} \put(4.05,2.65){-$\frac{g}{2}$}
\put(7.73,3){\line(0,1){0.12}} \put(7.63,2.65){$\frac{g}{2}$} 
\put(6,2){\line(1,0){0.12}}  \put(5.54,1.9){-$\frac{g^\prime}{2}$}
\put(6,4){\line(1,0){0.12}}  \put(5.65,3.9){$\frac{g^\prime}{2}$}
\put(4.27,2){\circle*{0.18}} \put(3.6,1.7){$\chi^-_{2L}$} 
\put(4.27,4){\circle*{0.18}} \put(3.6,4.2){$\chi^0_{2R}$}
\put(7.73,2){\circle*{0.18}} \put(7.9,1.7){$\chi^0_{1L}$}
\put(7.73,4){\circle*{0.18}} \put(7.9,4.2){$\chi^+_{1R}$}
\put(5.9,3){\circle*{0.18}}  \put(5.3,3.2){$\psi^0_{2L}$} 
\put(6.1,3){\circle*{0.18}}  \put(6.1,2.6){$\psi^0_{1R}$}
\put(6,1){\circle*{0.18}}    \put(5.3,0.9){$\psi^-_{2R}$} 
\put(6,5){\circle*{0.18}}    \put(5.3,4.9){$\psi^+_{1L}$}
\put(6.03,3.7){$\scriptstyle \theta_W$}
\end{picture} 
\vspace*{-15pt}

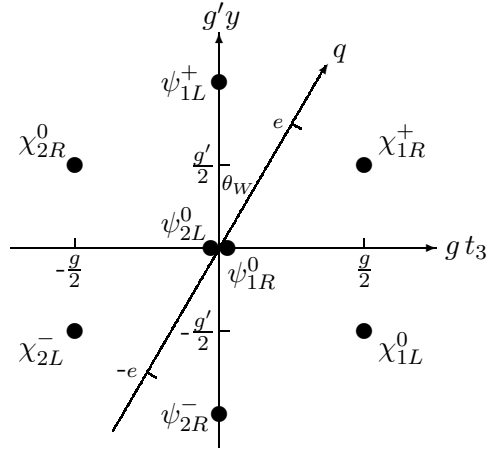
{\it Fig.1: The charges of the massless fermions forming leptons closely}$\qquad$\linebreak
\indent {\it approximate $\,SU(3)$ symmetry. A massive fermion consists of a singlet}$\qquad$\linebreak
\indent {\it state $\psi$ and a doublet state $\chi$ with opposite chirality, equal index} $i=1,2\qquad$\linebreak 
\indent {\it and equal electric charge}. \\

As Fig.1 shows, both the left-handed and the right-handed massless fermions forming leptons closely approximate a $SU(3)$ triplet and an uncharged singlet. An exact $SU(3)$ symmetry would require $\theta_W = 30^\circ$. Since $\theta_W$ slowly increases with increasing momentum transfer, there exists a momentum transfer for which the $SU(3)$ symmetry is exact. \\

For quarks, the $SU(3)$ symmetry is distorted due to their hypercharge $y=\frac{1}{6}$, $SU(3)$, however, requires $y=\pm \frac{1}{2}$. \\

Obviously, the elementary massless fermions form octets. The charges are quantised due to this symmetry. 

\pagebreak
3. In addition, a scalar field $\phi$ exists. The two massless fermions $\chi_{iL}$ and $\psi_{iR}$ interact via a Yukawa interaction, mediated by the scalar potential $\phi$. The Yukawa interaction is defined by the Lagrangian
\begin{equation} L_Y = - \sum_i C_i \left( \overline{\psi}_{iR} \, \phi^\dagger \chi_{iL} + \overline{\chi}_{iL} \phi \, \psi_{iR} \right) , \end{equation}
where $C_i$ is a charge. The Lagrangian $L_Y$ describes the reactions
\begin{equation} \begin{array}{c} \psi_{iR} + \phi \\ \chi_{iL} \end{array} \begin{array}{c} \rightarrow \\ \rightarrow \end{array} \begin{array}{c} \chi_{iL} \\ \psi_{iR} + \phi \, . \end{array} \end{equation}
These reactions conserve $U(1)$ charge and $t_3$. \\ 

The Lagrangian $L_Y$ implies that the scalar potential $\phi$ is a doublet. It is represented by
\begin{equation} \phi = \exp (- \mbox{i} \, \Theta ) \, {a \choose b} \, H \, , \end{equation}
where $H = H(x^\mu)$ is a scalar field with no internal degrees of freedom, representing the Higgs particle [4]. Due to the conservation of $U(1)$ charge by the reactions (19) and with the help of (11) the $U(1)$ charge of $\phi$ is
\begin{equation} g^\prime y_\phi = g^\prime y - g^\prime y_i = - g \, t_3 \tan \theta_W \, , \end{equation}
where the same $t_3$ is carried by $\chi_{iL}$ and $\phi$, since $t_3$ is conserved. Hence, the interaction transformation $U_\phi$ for an eigenstate of $T_3$ results in
\begin{eqnarray} U_\phi 
&=& \exp \Big(- \mbox{i} \int \left( \frac{g}{\sqrt{2}} (T_+ W_\mu^+ + T_- W_\mu^- ) + g\, t_3 W_\mu^3 + g^\prime y_\phi B_\mu \right) \mbox{d} x^\mu \Big) \nonumber\\[3pt] 
&=& \exp \Big(- \mbox{i} \int \left( \frac{g}{\sqrt{2}} (T_+ W_\mu^+ + T_- W_\mu^- ) + \frac{g\, t_3}{\cos \theta_W} \, Z_\mu \right) \mbox{d} x^\mu \Big) . 
\end{eqnarray}
The doublet $\phi$ does not carry electric charge and does not interact with the electromagnetic potential $A_\mu \,$, for the reason that $\chi_{iL}$ and $\psi_{iR}$ carry the same electric charge. \\

The parameters $a$ and $b$ of $\phi$ are identical with the parameters $a$ and $b$ of $\chi_{iL}$. The field $\exp (- \mbox{i} \, \Theta )$ implies a massless particle, the Goldstone boson, and represents the three remaining degrees of freedom of $\phi$, where
\begin{equation} \Theta = \Theta (x^\mu) = \frac{1}{\sqrt{2}} (T_+ \theta^+ + T_- \theta^- ) +  \frac{T_3}{\cos \theta_W} \, \theta^0 \, . \end{equation}  
The internal rotation angles $\theta^+ , \theta^- $ and $\theta^0$ correspond to the three internal degrees of freedom of $\phi$ represented by the gauge potentials of $U_\phi$. \\
 
The elementary fields and their interactions are thus defined. They imply the Lagrangian $\mathcal{L}$ for the electroweak interaction, consisting of the following terms
\begin{equation} \mathcal{L} = L_\psi + L_\chi + L_\phi + L_Y + L_W + L_B \, . \end{equation}
The terms of $\mathcal{L}$ are
\begin{eqnarray}
L_\psi  &=& \sum_i \, \overline{\psi}_{iR} \, \mbox{i} \, \gamma^\mu(\partial_\mu - M_\psi ) \psi_{iR} \\[4pt]
L_\chi &=&  \sum_i \, \overline{\chi}_{iL} \, \mbox{i} \, \gamma^\mu(\partial_\mu - M_\chi ) \chi_{iL} \\[-1pt]
L_\phi &=& \frac{1}{2} \left( \Big( (\partial_\mu - M_\phi ) \phi \Big)^\dagger \! \Big( (\partial^\mu - M_\phi ) \phi \Big) + \frac{m_H^2}{2}  \phi^\dagger \phi - \frac{m_H^2}{4 v^2} (\phi^\dagger \phi )^2 \right) \\[6pt] 
L_Y &=& \!\!\!\! - \sum_i C_i \left( \overline{\psi}_{iR} \, \phi^\dagger \chi_{iL} + \overline{\chi}_{iL} \phi \, \psi_{iR} \right) \\ 
L_W \! &=& \!\!\!\! - \frac{1}{4} \, \mathbf{W}_{\mu \nu} \mathbf{W}^{\mu \nu} \\[6pt]
L_B &=& \!\!\!\! - \frac{1}{4} \, F_{\mu \nu} F^{\mu \nu} \, ,
\end{eqnarray}
where $m_H$ is the mass of the Higgs particle, $v$ is a constant and the interaction matrices $M$ are due to (13) and (22)
\begin{eqnarray}
M_\psi \!\!\! &=& \!\!\! U_\psi^{-1} \partial_\mu U_\psi = - \mbox{i} \: q_i \left( A_\mu - \tan \theta_W Z_\mu \right) \nonumber\\[7pt]
M_\chi \!\!\! &=& \!\!\! U_\chi^{-1} \partial_\mu U_\chi = - \mbox{i} \left( \frac{g}{\sqrt{2}} (T_+ W_\mu^+ + T_- W_\mu^- ) + \frac{g \, t_3}{\cos \theta_W} \, Z_\mu + q_i \left( A_\mu - \tan \theta_W Z_\mu \right) \right) \nonumber\\[4pt] 
M_\phi \!\!\! &=& \!\!\! U_\phi^{-1} \partial_\mu U_\phi = - \mbox{i} \left( \frac{g}{\sqrt{2}} (T_+ W_\mu^+ + T_- W_\mu^- ) + \frac{g \, t_3}{\cos \theta_W} \, Z_\mu \right) .
\end{eqnarray} 
Hence, the complete Lagrangian of the standard model is defined. The elementary fields possess a well-defined internal symmetry and interact with massless gauge potentials, which is essential for a Yang-Mills interaction. \\

The Lagrangian $\mathcal{L}$ corresponds to the diagram displayed by Fig.2 showing the constituents of a massive fermion. \\

A massive fermion $\psi$ is formed by a left-handed massless fermion doublet state $\chi_L$ and a right-handed massless fermion singlet state $\psi_R$ exchanging a scalar doublet $\phi$. Thus the difference of the charges of $\chi_L$ and $\psi_R$ is transferred. The gauge potentials emitted by $\chi_L$ and $\psi_R$ denoted $M_\chi$ and $M_\psi$ interact with the scalar $\phi$ and finally obtain a longitudinal component and mass. By the $\phi$-exchange the constituents $\chi_L$ and $\psi_R$ are interchanged but the massive fermion $\psi$ remains unchanged.

\setlength{\unitlength}{1.1cm}
\begin{picture}(12,5.5) 
\put(2,2){\vector(1,0){8.1}} \put(2,4){\vector(1,0){8.1}}
\multiput(3,2)(0.222,0.222){16}{\qbezier(0,0)(0,0.11)(0.11,0.11)}
\multiput(3.111,2.111)(0.222,0.222){16}{\qbezier(0,0)(0.11,0)(0.11,0.11)}
\multiput(5,2)(0.222,0.222){16}{\qbezier(0,0)(0,0.11)(0.11,0.11)}
\multiput(5.111,2.111)(0.222,0.222){16}{\qbezier(0,0)(0.11,0)(0.11,0.11)}
\put(4,2){\circle*{0.11}} \put(6,2){\circle*{0.11}}
\put(8,2){\circle*{0.11}} \put(4,4){\circle*{0.11}}
\put(6,4){\circle*{0.11}} \put(8,4){\circle*{0.11}}
\put(3,2){\circle*{0.11}} \put(4,3){\circle*{0.11}}
\put(5,2){\circle*{0.11}} \put(6,3){\circle*{0.11}}
\put(1,2.9){$\psi$} \put(2.9,1.6){$\chi_L$} \put(4.9,1.6){$\psi_R$}
\put(2.9,4.2){$\psi_R$} \put(4.6,4.2){$\chi_L$} \put(4.1,2.7){$\phi$} 
\put(6.1,2.7){$\phi$} \put(6.4,5.1){$M_\chi$} \put(8.4,5.1){$M_\psi$} 
\linethickness{1pt}
\qbezier[20](4,2)(4,3)(4,4) \qbezier[20](6,2)(6,3)(6,4)
\qbezier[20](8,2)(8,3)(8,4)
\end{picture} 

\vspace*{-35pt}

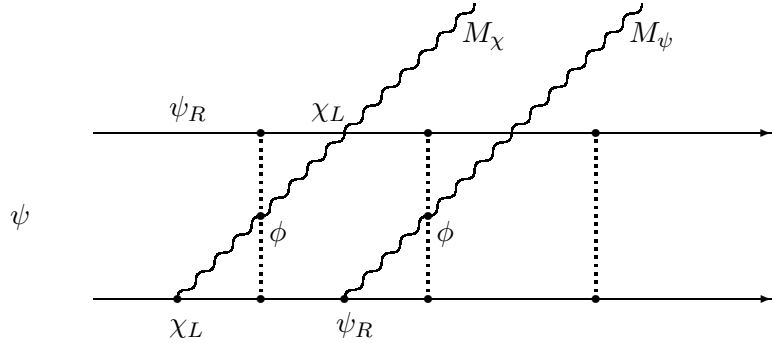
{\it Fig.2: The constituents of a massive fermion. A massive fermion $\psi$ consists}$\qquad$\linebreak
\indent {\it of a left-handed massless fermion doublet state $\chi_L$, a right-handed massless}$\qquad$\linebreak
\indent {\it fermion singlet state $\psi_R$ and a scalar doublet state $\phi$. By the exchange of}$\qquad$\linebreak
\indent {\it the scalar field $\phi$ the fermions $\chi_L$ and $\psi_R$ are interchanged. The gauge}$\qquad$\linebreak
\indent {\it potentials represented by the interaction matrices $M_\psi$ and $M_\chi$ interact}$\qquad$\linebreak
\indent {\it with $\phi$ and finally obtain a longitudinal component and mass.} \\[5pt]

The field $\phi$ introduced by (20) is consistent with the Lagrangian $L_\phi$. However, for this field the Lagrangian $\mathcal{L}$ does not appear acceptable for a theory of elementary particles, since for this field the Lagrangian $L_Y$ is not a Lorentz scalar and $L_\phi$ has a mass term with the wrong sign. Hence, $\phi$ corresponds to a particle, which exists only during an exchange. In order to obtain correct Lagrangians and a renormalisable theory [5], the field $\phi$ has to be redefined. \\ 

Inserting (4), (6) and (20) the Lagrangian $L_Y$ can be written as
\begin{equation} L_Y =  - \sum_i C_i \, H \, \overline{\psi}_i \left( \exp ( \mbox{i} \, \Theta ) \frac{1 - \gamma^5}{2} + \exp (- \mbox{i} \, \Theta ) \frac{1 + \gamma^5}{2} \right) \psi_i \, . \end{equation} 
A Lorentz scalar is only obtained if the $\gamma^5$ terms cancel. This is accomplished transforming $\phi$ into $\phi^\prime $ via
\begin{equation} \phi \rightarrow \phi^\prime = \exp ( \mbox{i} \, \Theta ) \, \phi = {a \choose b} H \, . \end{equation}
Thus, $\phi^\prime$ obtains the same internal coordinates $a$ and $b$ as $\chi_{iL}$. In order to leave the rest of the Lagrangian $\mathcal{L}$ invariant, the gauge potentials $W_\mu^+ , W_\mu^-$ and $Z_\mu$ are simultaneously transformed by a gauge transformation 
\begin{equation} W_\mu^+ \rightarrow W_\mu^+ + \frac{1}{g} \partial_\mu \theta^+ , \quad  W_\mu^- \rightarrow W_\mu^- + \frac{1}{g} \partial_\mu \theta^- , \quad  Z_\mu \rightarrow Z_\mu + \frac{1}{g} \partial_\mu \theta^0 . \end{equation}
The internal degrees of freedom of $\phi$ represented by the Goldstone boson are thus transferred to the gauge potentials, providing the gauge potentials with a third component. 

\pagebreak
Together with the third component the gauge potentials must acquire mass. This is accomplished by the particular structure of $L_\phi$. The Lagrangian $L_\phi$ consists of a kinetic energy density $T_\phi$ and a potential energy density $V_\phi$ 
\begin{equation} L_\phi = \frac{1}{2} \left( T_\phi - V_\phi \right) \, , \qquad \mbox{where} \qquad V_\phi = - \frac{m_H^2}{2} \, \phi^\dagger \phi + \frac{m_H^2}{4 v^2} (\phi^\dagger \phi)^2 . \end{equation}
The $(\phi^\dagger \phi)^2$ term represents a self-interaction. The vacuum state of $\phi$ is the field $\phi_0$ for which $V_\phi$ has its minimum. With the choice of the parameters of $L_\phi$ the minimum of $V_\phi \,$, for which
\begin{equation} 0 = \frac{\mbox{d} \, V_\phi}{\mbox{d} \, \phi} = - \frac{m_H^2}{2} \, \phi_0^\dagger + \frac{m_H^2}{2 v^2} \, \phi_0^\dagger (\phi_0^\dagger \phi_0) \, , \qquad \mbox{is at} \qquad \phi_0^\dagger \phi_0 = v^2 \, . \end{equation}
Thus, the vacuum state $\phi_0^\prime$ for the transformed field $\phi^\prime$ is obtained replacing $H$ by $v$
\begin{equation} \phi_0^\prime = {a \choose b} v \, , \end{equation}
and the field $\phi$ for a particle state $\phi^\prime$ created out of its vacuum state $\phi_0^\prime$ is given by
\begin{equation}
\phi = \phi_0^\prime + \phi^\prime = {a \choose b} \left(v + H \right) . \end{equation}
The Lagrangians $L_Y$ and $L_\phi$ imply that the general doublet field $\phi$ defined by (20) has to be redefined by this particular field $\phi$. Inserting the expression (38) for the field $\phi$ into $L_\phi$ and gives with
\begin{equation} (\partial_\mu - M_\phi )\phi \;
= \left( \begin{array}{c} \left( \partial_\mu + \mbox{i} \,\displaystyle \frac{g}{2} \frac{Z_\mu}{\cos \theta_W} \right) a + \mbox{i} \, g \displaystyle \frac{W_\mu^+}{\sqrt{2}} \, b \\  \mbox{i} \, g \displaystyle \frac{W_\mu^-}{\sqrt{2}} \, a+  \left( \partial_\mu - \mbox{i} \,\displaystyle \frac{g}{2} \frac{Z_\mu}{\cos \theta_W} \right) b \end{array} \right) \left(v + H \right) \end{equation} 

\begin{eqnarray}
L_\phi &=& \frac{1}{2} (\partial_\mu H)(\partial^\mu H) + \frac{g^2}{8} \left( 2 \, W_\mu^- W^{\mu +} + \frac{Z_\mu Z^\mu}{\cos^2 \theta_W} \right) \left(\phi_0^\dagger \phi_0 + 2 v H + H^2 \right) \nonumber\\
&+& \frac{m_H^2}{8 v^2} \left( 2 v^2 \left(v + H \right)^2 - \left(v + H \right)^4 \right) . \end{eqnarray}
This Lagrangian consists of two terms, $L_\phi = L_H + L_{\phi_0} \, , $ where $L_H$ is the Lagrangian for the observable Higgs particle 
\begin{eqnarray}
L_H &=& \frac{1}{2} (\partial_\mu H)(\partial^\mu H) - \frac{1}{2} m_H^2 H^2 + \frac{g^2}{8} \left(2 \, W_\mu^- W^{\mu +} + \frac{Z_\mu Z^\mu}{\cos^2 \theta_W} \right) \left(2 v H + H^2 \right) \nonumber\\
&+& \frac{m_H^2}{8 v^2} \left(v^4 - 4 v H^3 - H^4 \right)  \end{eqnarray}
with the mass $m_H \approx 125 \, \mbox{GeV}$ recently measured by the experiments ATLAS and CMS [6] at CERN. 

\pagebreak 
The remaining term $L_{\phi_0}$ describes the interaction of the gauge potentials with the vacuum state $\phi_0$. Since the Proca Lagrangian for massive vector fields, corresponding to (29), has the form
\begin{equation} L_W = \frac{1}{2} \, m_W^2 \mathbf{W}_\mu \mathbf{W}^\mu - \frac{1}{4} \mathbf{W}_{\mu \nu} \mathbf{W}^{\mu \nu} \, , \end{equation}
$L_{\phi_0}$ represents the required mass terms for the gauge potentials
\begin{eqnarray} L_{\phi_0}
&=& \frac{g^2}{8} \left(2 \, W_\mu^- W^{\mu +} + \frac{Z_\mu Z^\mu}{\cos^2 \theta_W}\right) \phi_0^\dagger \phi_0 \nonumber\\[5pt]
&=& \frac{g^2 v^2}{8} \left(W_\mu^{+ *} W^{\mu +} + W_\mu^{- *} W^{\mu -} + \frac{Z_\mu Z^\mu}{\cos^2 \theta_W} \right) \nonumber\\[5pt]
&=& \frac{1}{2} \, m_W^2 \left(W_\mu^{+ *} W^{\mu +} + W_\mu^{- *} W^{\mu -} + \frac{Z_\mu Z^\mu}{\cos^2 \theta_W} \right) . \end{eqnarray}
Thus, the gauge potentials $W_\mu^+$, $W_\mu^-$ and $Z_\mu$, which have obtained a longitudinal component, acquire mass, while $A_\mu$ remains massless. The gauge potential masses are 
\begin{equation} \, m_W = \frac{g}{2} \, v \quad \mbox{and} \quad m_Z = \frac{1}{\cos \theta_W} \, m_W \, .\end{equation} \\

The size of the constant $v$, which represents the electroweak mass scale, can be obtained via the Fermi coupling constant 
\begin{equation} G_F = \frac{\sqrt{2}}{8} \frac{g^2}{m_W^2} \approx 1.166 \cdot 10^{-5} \, \mbox{GeV}^{-2} \end{equation}
resulting in 
\begin{equation} v = \frac{1}{\sqrt{\sqrt{2} \, G_F}} \approx 246 \, \mbox{GeV} \, . \end{equation} \\  
 
The gauge potentials rearrange, forming the observable physical fields to which the mass terms give mass. This rearrangement of the gauge potentials is a global unitary transformation 
\begin{equation} \left( \begin{array}{c} \frac{1}{\sqrt{2}} \\ \frac{1}{\sqrt{2}} \\ 0 \\ 0 \end{array} \quad \begin{array}{c} \!\!\!\!\! - \frac{1}{\sqrt{2}} \, \mbox{i} \\ \frac{1}{\sqrt{2}} \, \mbox{i} \\ 0 \\ 0 \end{array} \begin{array}{c} 0 \\ 0 \\ \cos \theta_W \\ \sin \theta_W \end{array} \begin{array}{c} 0 \\ 0 \\ \!\!\!\! - \sin \theta_W \\ \cos \theta_W \end{array} \right) \left( \begin{array}{c} W_\mu^1 \\ W_\mu^2 \\ W_\mu^3 \\ B_\mu \end{array} \right) = \left( \begin{array}{c} W_\mu^+ \\ W_\mu^- \\ Z_\mu \\ A_\mu \end{array} \right) , \end{equation} \\
leaving the Lagrangian for the gauge potentials unchanged 
\begin{equation} L_W + L_B = L_{W^+} + L_{W^-} + L_Z + L_A \, . \end{equation} 

\pagebreak
By the Yukawa interaction of the three elementary fields $\chi_{iL}$, $\psi_{iR}$ and $\phi$ a massive fermion $\psi_i$ is formed. The two constituents $\chi_{iL}$ and $\psi_{iR}$ turn into the two components of $\psi_i$. The symmetry of the massive fermion state $\psi_i \,$, which is also the symmetry of its components, is the symmetry which $\chi_{iL}$ and $\psi_{iR}$ have in common. This is the symmetry of $\psi_{iR}$ as a comparison of $U_\chi$ with $U_\psi$ using (13) shows. Thus, $\psi_{iR}$ is the right-handed component of $\psi_i$ in agreement with its definition by (6). The corresponding left-handed component $\psi_{iL}$ of $\psi_i$ is due to (4) given by
\begin{equation} \chi_{iL} = {a \choose b} \psi_{iL} \, . \end{equation} 

Using this relation and inserting the redefined field $\phi$ given by (38) into the Lagrangian $L_Y$ results in a Lagrangian for a massive fermion $\psi_i$
\begin{eqnarray} L_Y \!
&=& \!\!\!\! - \sum_i C_i \left( \overline{\psi}_{iR} \, \phi^\dagger \chi_{iL} + \overline{\chi}_{iL} \phi \, \psi_{iR} \right) \nonumber\\ 
&=& \!\!\!\! - \sum_i C_i \left( \overline{\psi}_{iR} (v + H) (a^* , b^* ) {a \choose b} \psi_{iL} + \overline{\psi}_{iL} \, (a^* , b^* ) {a \choose b} (v + H) \psi_{iR} \right) \nonumber\\[5pt]
&=& \!\!\!\! - \sum_i C_i \left(v + H \right) \left( \overline{\psi}_{iR} \psi_{iL} + \overline{\psi}_{iL} \psi_{iR} \right) \nonumber\\[5pt]
&=& \!\!\!\! - \sum_i m_i \, \overline{\psi}_i \psi_i - \sum_i \frac{m_i}{v} \, \overline{\psi}_i H \, \psi_i \, . 
\end{eqnarray}
The first term is the mass term for a fermion $\psi_i$ with the mass
\begin{equation} m_i = C_i v \, , \end{equation}
and the second term describes the interaction of the Higgs particle $H$ with a massive fermion $\psi_i$. \\

As required by neutrino oscillations, also the neutrinos have mass. Hence, the charges $C_i$ are positive constants. \\

For the formation of a massive fermion $\psi_i$, the Lagrangian $L_\chi$ for the constituent $\chi_{iL}$ must be transformed into a Lagrangian for the left-handed component $\psi_{iL}$. Using the relation (49) gives
\begin{eqnarray} L_\chi
&=&  \sum_i \, \overline{\chi}_{iL} \, \mbox{i} \, \gamma^\mu(\partial_\mu - M_\chi ) \chi_{iL} \nonumber\\
&=& \sum_i \overline{\psi}_{iL} \, (a^* , b^* ) \, \mbox{i} \, \gamma^\mu (\partial_\mu - M_\chi ) \, {a \choose b} \, \psi_{iL} \nonumber\\
&=& \sum_i \overline{\psi}_{iL} \, (a^* , b^* ) {a \choose b}  \, \mbox{i} \, \gamma^\mu (\partial_\mu - M_\chi ) \, \psi_{iL} \nonumber\\[7pt] 
&=& \sum_i \overline{\psi}_{iL} \, \mbox{i} \, \gamma^\mu(\partial_\mu - M_\chi ) \, \psi_{iL} \, . 
\end{eqnarray} 

\pagebreak
Thus, the Lagrangians $L_\psi$ and $L_\chi$ can be expressed in terms of the massive fermion $\psi_i$ formed by the components $\psi_{iR}$ and $\psi_{iL}$. In addition, the interaction matrices $M$ given by (31) are inserted 
\begin{eqnarray}   
L_\psi &=& \sum_i \overline{\psi}_i \, \mbox{i} \, \gamma^\mu \frac{1+\gamma^5}{2}\Big( \partial_\mu + \mbox{i} \, q_i \left(A_\mu - \tan \theta_W Z_\mu \right) \Big) \, \psi_i \, , \nonumber\\[15pt]
L_\chi &=& \sum_i \overline{\psi}_i \, \mbox{i} \, \gamma^\mu \frac{1-\gamma^5}{2}\Big( \partial_\mu + \mbox{i} \, q_i \left(A_\mu - \tan \theta_W Z_\mu \right) \Big) \, \psi_i \nonumber\\  
&+& \sum_i \overline{\psi}_i \, \mbox{i} \, \gamma^\mu \frac{1-\gamma^5}{2} \, \mbox{i} \left(\frac{g}{\sqrt{2}} ( T_+ W_\mu^+ + T_- W_\mu^- ) + \frac{g \, t_3}{\cos \theta_W} \, Z_\mu \right) \psi_i \, . 
\end{eqnarray} 
 
Collecting the terms of the Lagrangian $\mathcal{L}$, using (50), (53), (41), (43) and (48), the complete Lagrangian $\mathcal{L}$ may be expressed in terms of the observable physical fields 
\begin{eqnarray} 
\mathcal{L} &=& \sum_i \overline{\psi}_i \left( \mbox{i} \, \gamma^\mu \partial_\mu - m_i - \frac{m_i}{v} H \right) \psi_i \nonumber\\
&-& \sum_i \overline{\psi}_i \, \gamma^\mu \frac{1- \gamma^5}{2} \left(\frac{g}{\sqrt{2}} ( T_+ W_\mu^+ + T_- W_\mu^- ) + \frac{g \, t_3}{\cos \theta_W} \, Z_\mu \right) \psi_i \nonumber\\[5pt]
&-& \sum_i \overline{\psi}_i \, \gamma^\mu q_i \left(A_\mu - \tan \theta_W Z_\mu \right) \psi_i \nonumber\\[2pt]
&+& \frac{1}{2} (\partial_\mu H)(\partial^\mu H) - \frac{1}{2} m_H^2 H^2 \nonumber\\[7pt]
&+& \frac{1}{8} \, g^2 \left( 2 \, W_\mu^- W^{\mu +} + \frac{Z_\mu Z^\mu}{\cos^2 \theta_W } \right)  \left(2 v H + H^2 \right) +  \frac{1}{8} \, \frac{m_H^2}{v^2} \left(v^4 - 4 v H^3 - H^4 \right) \nonumber\\[4pt]
&+& \frac{1}{2} \left( 2 m_W^2 W_\mu^- W^{\mu +} + m_Z^2 Z_\mu Z^\mu \right)\nonumber\\[8pt] 
&-& \frac{1}{4} \mathbf{W}_{\mu \nu} \mathbf{W}^{\mu \nu} - \frac{1}{4} F_{\mu \nu} F^{\mu \nu}  \, . 
\end{eqnarray}
This is the well established Lagrangian of the standard model [3]. Therefore, the original Lagrangian $\mathcal{L}$ introduced by (24) to (31) is valid. The choice of the elementary fields, their charges, their Lagrangians and their rearrangement, forming massive physical fields, as presented, is thus justified. \\
 
\newpage
{\bf References}

\vspace*{8pt}
\noindent
[1] \,S.L.\,Glashow, Nucl.Phys. {\bf 22} (1961) 579; \newline
\indent S.\,Weinberg, Phys.Rev.Lett. {\bf 19} (1967) 1264; \newline
\indent A.\,Salam, {\it Elementary Particle Theory}, ed. N.\,Swartholm, (Almquist and \newline
\indent Wiksell, Stockholm, 1968) \newline
[2] \,C.N.\,Yang and R.L.\,Mills, Phys.Rev. (1954) {\bf 69} 191 \newline
[3] \,Particle Data Group, {\it Review of Particle Physics}, Phys.Lett. {\bf B667} (2008) 125 \newline
\indent and references therein \newline
[4] \,F.\,Englert, R.\,Brout, Phys.Rev Lett. {\bf 13} (1964) 321 \newline  
\indent P.W.\,Higgs, Phys.Lett. {\bf 12} (1964) 132, Phys.Rev.Lett. {\bf 13} (1964) 508 \newline 
\indent P.W.\,Higgs, Phys.Rev. {\bf 145} (1964) 1156 \newline
\indent G.\,Guralnik, C.\,Hagen, T.W.B.\,Kibble, Phys.Rev.Lett. {\bf 13} (1964) 585 \newline
\indent T.W.B.\,Kibble, Phys.Rev. {\bf 155} (1967) 1554 \newline
[5] \,G.\,'t Hooft and M.\,Veltman Nucl. Phys. {\bf B44} (1972) 189 \newline
[6] \,CMS Collaboration, S.\,Chatrchyan el al. Phys.Lett. {\bf B710} (2012) 26  \newline
\indent ATLAS Collaboration, G.\,Aad et al. Phys.Lett. {\bf B710} (2012) 383 \newline 
[7] \,L.H.\,Ryder, {\it Quantum Field Theory} (Cambridge University Press, 1985, 1996) \newline
\indent and references therein.

\newpage

{\bf Appendix}\\ 

Basics of Yang-Mills theory [7] \\

{\it One internal dimension}

A complex particle field $\Psi = \Psi \Big(\exp (- \mbox{i} \, \varphi) \Big)$ undergoes a $U(1)$ interaction, transforming the field $\Psi$ into the field $\Psi^\prime$
\begin{equation} \Psi \rightarrow \Psi^\prime = U_0 \Psi \, ,\end{equation}
where $U_0$ is a unitary transformation given by 
\begin{equation} U_0 = \exp \Big(- \mbox{i} \, \int_{x^\mu}^{x^{\prime \mu}} g_0 B_\mu \, \mbox{d}x^\mu \Big) .\end{equation}
When a complex field carrying the $U(1)$ charge $g_0$ moves from space-time point $x^\mu$ to space-time point $x^{\prime \mu}$ interacting with a real, massless gauge potential $B_\mu$, the internal variable $\varphi$ is changed by $\Delta \varphi$. Simultaneously, the potential $B_\mu$ undergoes a gauge transformation
\begin{equation} B_\mu \rightarrow B_\mu  - \frac{1}{g_0} \partial_\mu (\Delta \varphi) \end{equation} so that the total Lagrangian remains invariant under the interaction. The momentum operator for the free field $\mbox{i} \partial_\mu$ due to the interaction gets the eigenvalue $p_\mu + \mbox{i} \, U_0^{-1} \partial_\mu U_0$. Therefore, the momentum operator has to be modified by the replacement
\begin{equation} \partial_\mu \; \rightarrow \; \partial_\mu - U_0^{-1} \partial_\mu U_0 \, .\end{equation}
The field strength $F_{\mu \nu}$ created by the $U(1)$ potential $B_\mu$ is
\begin{equation} F_{\mu \nu} = \partial_\mu B_\nu - \partial_\nu B_\mu \, . \end{equation}
It can be shown that the total $U(1)$ charge is conserved. \\

{\it Three internal dimensions}

A particle field $\Psi = \Psi(\mathbf{r})$, being equivalent to a real 3-vector $\mathbf{r}$, undergoes a $SO(3)$ interaction. By the interaction, the vector $\mathbf{r}$ is rotated, and $\Psi$ is transformed into $\Psi^\prime$ via a unitary transformation $U_R$
\begin{equation} \Psi^\prime = U_R \Psi(\mathbf{r}) = \Psi(R \mathbf{r}) \, ,\end{equation}
where the rotation $R$ is given by
\begin{equation} R = \exp \Big( \mbox{i} \int g \sum_\ell J_\ell W_\mu^\ell \, \mbox{d} x^\mu \Big) \end{equation}
and the rotation axes are denoted by $\ell$. The generators of the rotations $J_\ell$ are
\begin{equation} J_1 = \left( \begin{array}{ccr} 0&0&0\\0&0&\!\!\!\!-\mbox{i}\\0&\mbox{i}&0 \end{array} \right) , \quad
J_2 = \left( \begin{array}{rcc} 0&0&\mbox{i}\\0&0&0\\-\mbox{i}&0&0 \end{array} \right) , \quad
J_3 = \left( \begin{array}{crc} 0&\!\!\!\!-\mbox{i}&0\\ \mbox{i}&0&0\\0&0&0 \end{array} \right) . \end{equation}
The charge $g$ is the universal $SO(3)$ charge, and the gauge potentials $W_\mu^\ell$ form a real 3-vector $\mathbf{W}_\mu$ of potentials. Moreover, the field strength $\mathbf{W}_{\mu \nu}$ created by the potential $\mathbf{W}_\mu $ is, after some algebra
\begin{equation} \mathbf{W}_{\mu \nu} = \partial_\mu \mathbf{W}_\nu -\partial_\nu \mathbf{W}_\mu + g \, \mathbf{W}_\mu \times \mathbf{W}_\nu \, . \end{equation} \\

{\it Four internal dimensions}

A doublet of particle fields is represented by $\Psi = \Psi_0 \displaystyle{a \choose b}$, where $a$ and $b$ are complex numbers with $a^* a + b^* b = 1$. \\
A complex 2-spinor $\displaystyle{a \choose b}$ forms a real 4-vector $r_\nu = (a^* , b^* ) \, \sigma_\nu \, \displaystyle{a \choose b} $
with the Pauli matrices $\sigma_\nu$
\begin{equation} 
\sigma_0 = \left( \begin{array}{cc} 1 & 0 \\ 0 & 1 \end{array} \right) , \quad
\sigma_1 = \left( \begin{array}{cc} 0 & 1 \\ 1 & 0 \end{array} \right) , \quad 
\sigma_2 = \left( \begin{array}{cr} 0 &\!\!\!-\mbox{i}\\ \mbox{i} & 0 \end{array} \right) , \quad 
\sigma_3 = \left( \begin{array}{cr} 1 & 0 \\ 0 &\!\!\!-1 \end{array} \right) . \end{equation}
Since the doublet is equivalent to a 3-vector, it undergoes an interaction transformation $U_R$, resulting from a rotation $R$ of the 3-vector part $\mathbf{r}$ of $r_\nu$ 
\begin{equation} (a^* , b^* ) U_R^\dagger \, \sigma_\nu \, U_R {a \choose b} = R \left( (a^* , b^* ) \, \sigma_\nu {a \choose b} \right), \qquad \nu = 1,2,3 \,   . \end{equation}
This equation is solved by
\begin{equation} U_R = \exp \Big( \pm \, \mbox{i} \int g \sum_\ell \frac{\sigma_\ell}{2} W_\mu^\ell \, \mbox{d}x^\mu \Big) \, . \end{equation} 
The interaction transformation $U_R$ of the doublet is hence the $SU(2)$ transformation, where the charge $g$ is the universal $SO(3)$ charge. \\

To the $SU(2)$ interaction, a $U(1)$ interaction can be added, using the not yet used  linearly independent $\sigma$ matrix $\sigma_0$ as generator. In this case, the total interaction $U$ of a doublet is the $SU(2) \times U(1)$ interaction, where $B_\mu$ is orthogonal to the $W_\mu$ potentials
\begin{eqnarray} U &=& \exp \Big( -\mbox{i} \int \left( g \sum_\ell \frac{\sigma_\ell}{2} W_\mu^\ell + g^\prime y \, \sigma_0 B_\mu \right) \mbox{d}x^\mu \Big) \nonumber\\[6pt] 
&=& \exp \Big( -\mbox{i} \int \left( \begin{array}{cc} g^\prime y \, B_\mu + \displaystyle\frac{g}{2} W_\mu^3 & \displaystyle\frac{g}{2} (W_\mu^1 - \mbox{i} \, W_\mu^2 ) \\[16pt] \displaystyle\frac{g}{2} (W_\mu^1 + \mbox{i} \, W_\mu^2 ) & g^\prime y \, B_\mu - \displaystyle\frac{g}{2} W_\mu^3 \end{array} \right) \mbox{d} x^\mu \Big) \, . \end{eqnarray}
For convenience the $U(1)$ charge $g^\prime y$ of the doublet is written as two factors, where $g^\prime$ serves as the $U(1)$ unit charge, and $y$ is called the weak hypercharge. 

The potentials implied by the $SU(2)$ interaction form an isospinor $\mathcal{W}$ containing complex, charged fields 
\begin{equation} \mathcal{W} = \left( \begin{array}{c} W^+ \\[6pt] W^0 \\[6pt] W^- \end{array} \right)= \left( \begin{array}{c} \displaystyle\frac{1}{\sqrt{2}} (W^1 - \mbox{i} \, W^2 ) \\ W^3 \\ \displaystyle\frac{1}{\sqrt{2}} (W^1 + \mbox{i} \, W^2 ) \end{array} \right) , \qquad \mathcal{W}^\dagger \mathcal{W} = \mathbf{W}^2 \, . \end{equation}
The $SU(2) \times U(1)$ interaction transformation of a doublet can be rewritten in terms of the charged potentials introducing the operators 
\begin{equation} T_\pm = \frac{1}{2} (\sigma_1 \pm \mbox{i} \, \sigma_2) \, , \qquad T_3 = \frac{1}{2} \sigma_3 \, , \qquad \mbox{and} \qquad Y = y \, \sigma_0 \, , \end{equation}
resulting in  
\begin{equation} U = \exp \Big(-\mbox{i} \int \left( g \left( \frac{1}{\sqrt{2}} ( T_+ W_\mu^+ + T_- W_\mu^- ) + T_3 W_\mu^3 \right) + g^\prime Y B_\mu \right) \mbox{d}x^\mu \Big) \, . \end{equation} 
By an interaction, the doublet $\Psi $ is transformed into $\Psi^\prime $ via 
\begin{equation} \Psi^\prime = U \, \Psi \, .\end{equation}
 
\end{document}